\documentclass[12pt,a4paper]{article}
\usepackage{amssymb}
\usepackage{amsmath}
\usepackage{amsfonts}

\begin{document}

\begin{center}
\huge{\bf The vacuum electromagnetic fields and the Schr\"{o}dinger picture}\\[0.3cm]
\end{center}

\vspace{0.5cm}

\begin{center}
A. J. Faria, H. M. Fran\c{c}a\footnote{e-mail: hfranca@if.usp.br}, G. G. Gomes and R. C. Sponchiado\\[0.3cm]
\end{center}

\begin{center}
{\em Instituto de F\'{i}sica, Universidade de S\~{a}o Paulo \\
C.P. 66318, 05315-970 S\~{a}o Paulo, SP, Brazil}
\end{center}

\vspace{1cm}

\begin{abstract}

Several authors have used the {\it Heisenberg picture} to show that
the atomic transitions, the stability of the ground state and the
position-momentum commutation relation $[x,p]=i\hbar$, can only be
explained by introducing radiation reaction and vacuum
electromagnetic fluctuation forces. Here we consider the simple case
of a nonrelativistic charged harmonic oscillator, in one dimension,
to investigate how to take into account the radiation reaction and
vacuum fluctuation forces within the {\it Schr\"{o}dinger picture}.
We consider the effects of both classical {\it zero-point} and {\it thermal}
electromagnetic vacuum fields. We show that the zero-point
electromagnetic fluctuations are dynamically related to
the momentum operator $p=-i \hbar\partial/\partial x$ used in the
Schr\"{o}dinger picture. Consequently, the introduction of the {\it
zero-point} electromagnetic fields in the vector potential $A_x(t)$
used in the Schr\"{o}dinger equation, generates ``double counting'',
as was shown recently by A.J. Faria et al. (Physics Letters A
\underline{305} (2002) 322). We explain, in details, how to avoid the
``double counting'' by introducing only the radiation reaction and the 
{\it thermal} electromagnetic fields into the Schr\"{o}dinger equation.
\end{abstract}




\section*{1. Introduction}

The question concerning the equivalence between the  Schr\"{o}dinger
and the Heisenberg pictures of quantum mechanics was raised a long
time ago by P.A.M. Dirac \cite{Dirac}. Recently, A.J. Faria et al.
\cite{Faria} have addressed the problem of the equivalence between
the two pictures. In order to clearly explain this problem we start
by indicating the importance of the radiation reaction and the
vacuum zero-point electromagnetic fields to the understanding of the
atomic transitions, and the atomic stability, using the Heisenberg
picture and quantum electrodynamics.

Consider a physical system like the hydrogen atom. Its Hamiltonian is
\begin{equation}
H_S = \frac{\vec{p\,}^2}{2m} - \frac{e^2}{r}\;,
\label{1}
\end{equation}
and the atomic states are such that
\begin{equation}
H_S|{\rm{vac}},a\rangle=\epsilon_a|{\rm{vac}},a\rangle\; ,
\label{2}
\end{equation}
where $\epsilon_a$ is the energy of the atom and $|{\rm{vac}},a\rangle \equiv |{\rm{vac}}
\rangle |a\rangle$ denotes the state in which the atom is in the stationary state
  $|a\rangle$, and the field is in its vacuum state
  $|{\rm{vac}}\rangle$ of no photons. Considering the above system,
  Dalibard, Dupont-Roc and Cohen-Tannoudji \cite{Dalibard} have
  discussed the role of the {\it vacuum zero-point
  fluctuations and the radiation reaction} forces, {\it with the identification of their respective
contributions}, in the domain of the atomic transitions with
emission of electromagnetic radiation. Considering the conceptual
importance of this paper we summarize their main conclusion.

Using a perturbative calculation based on the {\it Heisenberg
picture}, Dalibard et al. concluded that the
variation with time of the energy of the system is such that
\begin{eqnarray}
\langle {\rm{vac}},a|\frac{dH_S}{d\,t}|{\rm{vac}},a\rangle = -\frac{2}{3}
\frac{e^2}{c^3}\langle a| (\ddot{\vec{r}})^2 |a\rangle +
\nonumber \\  +
\frac{2}{3}\frac{e^2}{c^3}
\left[\sum_{b\,(\epsilon_{b}>\epsilon_{a})}\langle a| \ddot{\vec{r}}
|b\rangle \cdot \langle b| \ddot{\vec{r}} |a\rangle-\sum_{b\,(\epsilon_{b}<\epsilon_{a})} \langle a| \ddot{\vec{r}}
|b\rangle \cdot \langle b| \ddot{\vec{r}} |a\rangle \right]\; .
\label{3}
\end{eqnarray}
The first term in (\ref{3}) is the contribution of radiation reaction
whereas the second, and the third terms, are the contributions of
the vacuum fluctuation forces.
It is straightforward to show that (\ref{3}) can be written as
\begin{equation}
\langle {\rm{vac}},a|\frac{dH_S}{d\,t}|{\rm{vac}},a\rangle= - \frac{4}{3}
\frac{e^2}{c^3} \sum_{b(\epsilon_b<\epsilon_a)} \langle a|
\ddot{\vec{r}} |b\rangle \cdot \langle b| \ddot{\vec{r}} |a\rangle\; .
\label{4}
\end{equation}
We note that, if self reaction was alone (see the first term in
(\ref{3})), the atomic ground state would not be stable, since the
square of the acceleration has a non zero average value in such a
state. Moreover, such a result is extremely simple and exactly
coincides with what is found in classical radiation theory
\cite{Dalibard}. The complete result (see equation
(\ref{4})), which includes the vacuum forces, is even more
satisfactory because the electron in the vacuum can only lose energy
by cascading downwards to lower energy levels. The ground state
cannot be stable in the absence of vacuum fluctuations which exactly
balance the energy loss due to self reaction \cite {Franca}. In other words, if self
reaction was alone, the ground state would collapse and the atomic
commutation relation $[x,p]=i\hbar$ would not hold \cite{Dalibard}.
As stated in reference \cite{Dalibard}, ``all self reaction
effects, which are independent of $\hbar$, are strictly identical to
those derived from classical radiation theory. All zero-point vacuum
fluctuation effects, which are proportional to $\hbar$ can be interpreted
by considering the vibration of the electron induced by a random
field having a spectral power density equal to $\hbar\omega/2$
per mode''. Therefore, in several situations, the zero-point and
thermal vacuum fields can be successfully replaced by {\em classical} random fields
\cite{Franca, Marshall,Boyer, Milonni76, Pena}, so that the electric and magnetic fields can
be considered as fluctuating sources of energy.

In order to clarify the features of the interaction between the atom
and the vacuum fluctuating fields, we study the statistical
properties of a charged harmonic oscillator interacting with vacuum
fields, using the {\it Schr\"{o}dinger picture}. This is discussed
in section 2. We consider separately the effects of each kind of
fluctuating field. The effects of the {\it zero-point} radiation and
the radiation reaction are analized within the subsection 2a. The
effects of the {\it thermal} radiation and the radiation reaction
are studied within the subsection 2b. Conclusions are presented in
the section 3.


\section*{2. Charged harmonic oscillator according to the Schr\"{o}dinger picture}

For clarity reason and in order to simplify the calculations, the
classical vacuum electric fields to be considered here are the
random {\it zero-point} and {\it thermal} electric fields of
Stochastic Electrodynamics (SED). An excellent review of SED is
given in the book by de la Pe\~{n}a and Cetto \cite{Pena}.

We shall assume that the motion of the charged oscillator is non
relativistic ($mc^2 >> \hbar \omega_0$) so that the dipole
approximation will be used \cite{Boyer}. We shall see that this
approximation is consistent with the calculations presented in the
subsection 2.1 and 2.2. Following the notation of Boyer
\cite{Boyer}, the $x$ component of the zero-point
electric field, acting on the bounded charge moving close to the origin of the coordinate
system, is
\begin{equation}
E_x(\vec{r},t) = \sum_{\lambda=1}^{2} \int d^3k
\epsilon_x(\vec{k},\lambda) \frac{\sqrt{\hbar\omega/2}}{2\pi} \left[
e^{i\theta(\vec{k},\lambda)} e^{-i\omega t}e^{i\vec{k}\cdot\vec{r}} +
c.c. \right] \;.
\end{equation}

In the long wavelength approximation, one can write this expression as
a function of the time $t$ only, that is, 
\begin{equation}
 E_0(t) \simeq \sum_{\lambda=1}^{2} \int d^3k \epsilon_x(\vec{k},\lambda) 
\frac{\sqrt{\hbar\omega/2}}{2\pi} \left[e^{i\theta(\vec{k},\lambda)}
  e^{-i\omega t} + c.c. \right] \; , 
\label{E0}
\end{equation}
because the large values of $\mid \vec{k} \mid$ will not contribute
to the motion of a charge bounded by a harmonic force (frequency
$\omega_0$). This will be very clear in the next section (see also
T.H. Boyer \cite{Boyer}).

In (\ref{E0}), $\theta(\vec{k},\lambda)$ are random phases
statistically independent and uniformly distributed in the interval
$[0,2\pi]$, $\vec{k}$ is the wave vector such that $|\vec{k}| =
\omega/c$, and $\epsilon_x(\vec{k},\lambda)$ is the polarization
vector projected in the $x$ axis, with $\lambda = 1 \, , \, 2$.
Notice that spectral density of the zero-point radiation is such
that \cite{Marshall,Boyer}
\begin{equation}
\rho_0(\omega) = \frac{\hbar\omega^3}{2\pi^2 c^3}\;.
\label{espectro}
\end{equation}
The thermal electric field $E_T(t)$ is also random and is, by assumption, statistically
independent from $E_0(t)$. It can be written in a similar manner, namely
\begin{equation}
E_T(t) \simeq \sum_{\lambda=1}^{2} \int d^3k
\epsilon_x(\vec{k},\lambda) \frac{h(\omega,T)}{2\pi}[
e^{i\theta(\vec{k},\lambda)} e^{-i\omega t} + c.c.]\;, 
\label{ET}
\end{equation}
where $T$ is the absolute temperature and the function $h(\omega,T)$
is given by
\begin{equation}
h(\omega,T) =
\sqrt{\frac{\hbar\omega}{2}[\;\coth(\frac{\hbar\omega}{2kT}) - 1\;]}\;.
\label{hT}
\end{equation}
Notice that $h(\omega,T)=0$ if $T=0$. The spectral density of the
thermal radiation is such that
\begin{equation}
\rho_T(\omega) = \frac{\hbar\omega^3}{\pi^2c^3}\left(\frac{1}{e^{\hbar\omega/kT}-1}\right)\;.
\label{especterm}
\end{equation}

The system we shall study is a charged harmonic oscillator with
natural frequency $\omega_0$ and mass $m$ ($mc^2 >> \hbar
\omega_0$), already considered in a previous work \cite{Faria}. We
shall consider firstly the effects of the zero-point electric field
$E_0(t)$, given in (\ref{E0}), and the radiation reaction force. The
electric field associated with the radiation reaction will be
denoted by $E_{RR}(t)$ and will be obtained later. The above fields
will be introduced into the Schr\"{o}dinger equation through the
vector potential $A_x(t)$ such that
\begin{equation}
-\frac{1}{c}\frac{\partial A_x}{\partial t} = E_0(t) + E_{RR}(t) \;,
\end{equation}
in the case of {\it zero-point} radiation, or
\begin{equation}
-\frac{1}{c}\frac{\partial A_x}{\partial t} = E_T(t) + E_{RR}(t) \;,
\end{equation}
in the case of {\it thermal} radiation.


\subsection*{2.1) The effects of the zero-point field and the
  radiation reaction in the  Schr\"{o}dinger equation}

For reader convenience we shall obtain, in what follows, an exact
solution of the Schr\"{o}dinger equation by using the same method
already presented in the reference \cite{Faria}. By considering the
dipole (or long wavelength) approximation the one dimensional
Schr\"{o}dinger equation takes the form \cite{Faria}
\begin{equation}
i \hbar \frac{\partial \psi(x,t)}{\partial\, t} = \left[
\frac{1}{2m} \left( -i\hbar \frac{\partial}{\partial x} -
\frac{e}{c}A_x(t) \right)^2 + \frac{m \omega_{0}^{2} x^2}{2}
\right] \psi(x,t)\; ,
\label{Sch}
\end{equation}
where $A_x(t)$ is $x$ component of the vector potential acting on the
charged particle. At this point the exact analytical form of $A_x(t)$
is not known, because the radiation reaction field $E_{RR}(t)$ was not
determined. For the moment we shall simply assume that $A_x(t)$ is a
c-number that varies with $t$ and is {\em independent} of $x$. It
should be noticed that this assumption is valid provided that $mc^2
\gg \hbar \omega_0$.

The time independent Schr\"{o}dinger equation has a ground state
solution $\phi_0(x)$ such that
\begin{equation}
\phi_0(x) = \left( \frac{m \omega_0}{\pi \hbar}
\right)^{\frac{1}{4}} \exp{\left( - \frac{m \omega_0 x^2}{2 \hbar}
\right)}\; .
\label{15}
\end{equation}
Moreover, we see that
\begin{equation}
\int_{-\infty}^{\infty} dx\, \phi_{0}^{2}(x)
x^2 = \frac{\hbar}{2 m \omega_0}\; .
\label{16}
\end{equation}
The time dependent equation (\ref{Sch}) has an exact solution that
can be written as
\begin{equation}
\psi(x,t) = \phi_{0} \left( x-q_c(t) \right) \exp{ \left\{
\frac{i}{\hbar} \left[ \left( p_c(t) +\frac{e}{c}A_x(t) \right) x - g(t)
\right] \right\} }\; ,
\label{17}
\end{equation}
where the functions $q_c(t)$, $p_c(t)$ and $g(t)$ are unknown
c-numbers that will be determined by the substitution of (\ref{17})
into (\ref{Sch}). This is an old procedure, introduced by
Schr\"{o}dinger (1926) in a famous paper entitled {\it ``The
Continuous Transition from Micro to Macro-Mechanics''} (see
reference \cite{Schrodinger}, pg. 41). With the above substitution,
we get the following equations \cite{Husimi}:
\begin{equation}
p_c(t) = m \dot{q}_c(t)\; ,
\label{18}
\end{equation}
and
\begin{equation}
\dot{p}_c(t) = - m\omega_{0}^2 q_c(t) - \frac{e}{c}
\frac{\partial A_x(t)}{\partial t}\; .
\label{19}
\end{equation}
We also obtain the equation $2\dot{g}(t) = \hbar\omega_0 +
m\dot{q}_c^2(t) - m\omega_0^2 q_c^2(t)$, which solution can be written as
\begin{equation}
g(t) = \frac{\hbar\omega_0 t}{2} + \frac{m}{2}
\int_0^t dt' \left( \dot{q}_c^2(t') - \omega_0^2 q_c^2(t') \right)\; .
\label{20}
\end{equation}
One can combine (\ref{18}) and (\ref{19}) to obtain the differential
equation
\begin{equation}
m \ddot{q}_c(t) = -m\omega_0^2 q_c(t) + e E_x(t)\; ,
\label{21}
\end{equation}
where we have used the fact that $cE_x(t) = -\partial A_x(t)/\partial
t$. Notice that, by assumption, every term in (\ref{21}) is a
c-number. According to our definition, the total electric field will
be given by
\begin{equation}
E_x(t) = E_0(t) + E_{RR}(t)\; ,
\label{22}
\end{equation}
where $E_0(t)$ (see equation (\ref{E0}))
is the classical zero-point field and $E_{RR}(t)$ is the classical
radiation reaction field (the particle is {\em charged}, therefore the
radiation reaction field must contribute to $E_x(t)$).

The correct expression for the classical radiation reaction force
$eE_{RR}(t)$ is more difficult to obtain because, according to the
Schr\"{o}dinger picture, the charged particle does not have a precise
location. One can only say that
\begin{equation}
|\psi(x,t)|^2 =
\left( \frac{m\omega_0}{\pi\hbar} \right)^{\frac{1}{2}}
\exp{\left[ - \frac{m \omega_0 (x - q_c(t))^2}{\hbar} \right]}\; ,
\label{23}
\end{equation}
is the time dependent probability density. Notice that, in order to obtain
(\ref{23}), one must solve (\ref{21}) which depends on the still
undefined radiation reaction force $eE_{RR}(t)$. This force, however,
can be precisely defined in the case of large mass, so that
$mc^2 \gg \hbar\omega_0$. In this case, one can safely
consider that
\begin{equation}
eE_{RR}(t) \simeq \frac{2e^2}{3c^3} \stackrel{...}{q}_c(t)\;,
\label{24}
\end{equation}
is a good approximation because the Gaussian (\ref{23}) is so
narrow that the harmonically bound particle has a {\it trajectory}.
Based on these considerations we conclude that the
expression (\ref{24}) is valid in the case $mc^2 \gg \hbar \omega_0$,
which is consistent with the {\em long wavelength}
approximation. Therefore the equation (\ref{21}) can be written as
\begin{equation}
\ddot{q}_c(t) + \omega_0^2 q_c(t) \simeq \frac{e}{m} E_0(t) +
\frac{2e^2}{3mc^3} \stackrel{...}{q}_c(t)\; ,
\label{25}
\end{equation}
where $E_0(t)$ is given by (\ref{E0}). The last term in (\ref{25}) is
responsible for the decay of the excited states of the
oscillator.

The stationary solution of the equation (\ref{25}) is given by \cite{Marshall,Boyer}
\begin{equation}
q_c(t) = \frac{e}{m} \sum_{\lambda=1}^{2} \int d^3k
\epsilon_x(\vec{k},\lambda) \frac{\sqrt{\hbar\omega/2}}{2\pi} \left[
\frac{e^{i\theta(\vec{k},\lambda)} e^{-i\omega t}}{\omega_0^2 -
\omega^2 - i\frac{2e^2}{3mc^3} \omega^3} + c.c. \right]\; ,
\label{stat}
\end{equation}
which is a random real function of the time. This stationary
solution is obtained only if $e \neq 0$. If $e = 0$ the equation
(\ref{25}) will lead to oscillatory (non dissipative) coherent states
of the harmonic oscillator.

According to the Max Born statistical interpretation of the wave function $\psi(x,t)$, the
expectation value of $x^2$ is given by
\begin{equation}
\overline{x^2(t)} =
\int_{-\infty}^{\infty}dx |\psi(x,t)|^2\, x^2 .
\label{26}
\end{equation}
Taking into account the expressions (\ref{17}) and (\ref{15}), one
can show that
\begin{eqnarray}
\overline{x^2(t)} & = &
\int_{-\infty}^{\infty}dx \phi_{0}^2(x-q_c(t))\,
\left[\left(x-q_c(t)\right)^2+q_c^2(t) \right]
\nonumber \\ & = &
\frac{\hbar}{2m\omega_0} + q_c^2(t) .
\label{27}
\end{eqnarray}
We recall that $q_c^2(t)$ depends on the random phases
$\theta(\vec{k},\lambda)$.

From the result (\ref{27}) we can calculate the mean square value of
the particle position. This quantity is obtained by averaging over
the random phases present in (\ref{stat}).

The average over the random variables (indicated by the symbol
$\langle \, \rangle$) is such that \cite{Boyer}
\begin{equation}
\begin{array}{c}
\langle e^{i\theta(\vec{k},\lambda)} e^{i\theta(\vec{k}',\lambda')} \rangle =
\langle e^{-i\theta(\vec{k},\lambda)} e^{-i\theta(\vec{k}',\lambda')}
\rangle = 0 \; ,
\\
\langle e^{i\theta(\vec{k},\lambda)} e^{-i\theta(\vec{k}',\lambda')} \rangle =
\delta_{\lambda\lambda'} \delta^3(\vec{k}-\vec{k}')\; .
\end{array}
\label{ran}
\end{equation}
Hence, applying the random average to the expression (\ref{27}), we
obtain
\begin{equation}
\langle \overline{x^2} \rangle = \frac{\hbar}{2m \omega_0} +
\langle q_c^2(t) \rangle .
\label{28}
\end{equation}

Using the stationary solution (\ref{stat}), the average of $q_c^2(t)$ over the
random phases is such that \cite{Marshall,Boyer}
\begin{equation}
\langle q_c^2(t) \rangle =
\frac{2 e^2}{3\pi m^2c^3} \int_0^{\infty}
d\omega \frac{\hbar\omega^3}{ ( \omega^2 - \omega_0^2 )^2 +
\left( \frac{2e^2}{3mc^3} \right)^2 \omega^6}\;.
\label{28b}
\end{equation}
Since $(\frac{2}{3}\frac{e^2}{\hbar c}\frac{\hbar \omega_0}{m
c^2})^2 \ll 1$, the integrand of (\ref{28b}) has a very sharp peak
at $\omega \approx \omega_0$. Therefore, this integral can be
approximated by (see \cite{Boyer} and also the Appendix A of the
reference \cite{Milonni})
\begin{equation}
\langle q_c^2(t) \rangle \simeq
\frac{2\hbar\omega_0^3 e^2}{3\pi m^2c^3} \int_{0}^{\infty}
\frac{d\omega}{ 4 \omega_0^2 ( \omega - \omega_0 )^2 +
\left( \frac{2e^2\omega_0^3}{3mc^3} \right)^2 }.
\label{28c}
\end{equation}

This expression can be cast in a more simple form, namely
\begin{equation}
\langle q_c^2(t) \rangle =
\frac{\hbar \gamma}{4\pi m \omega_0} \int_{0}^{\infty}
\frac{d\omega}{ ( \omega - \omega_0 )^2 + (\gamma / 2)^2 },
\label{29}
\end{equation}
where $\gamma \equiv \frac{2}{3}\frac{e^2\omega_0^2}{mc^3}$, and
$\gamma/\omega_0 \ll 1$. This is a standard integral and the result
is
\begin{equation}
\langle q_c^2(t) \rangle =
\frac{\hbar}{2\pi m \omega_0}[\;\frac{\pi}{2} + \arctan(\frac{2
\omega_0}{\gamma})\;],
\label{31}
\end{equation}
showing that $\langle q_c^2(t) \rangle$ is charge dependent because
$\frac{\gamma}{\omega_0}=\frac{2e^2\omega_0}{3mc^3}$. An expansion of
(\ref{31}) in powers of the small constant $\gamma/\omega_0$ gives
\begin{equation}
\langle q_c^2(t) \rangle =
\frac{\hbar}{2 m \omega_0}[\;1-
\frac{1}{\pi}(\frac{\gamma}{2\omega_0}) +
\frac{1}{3\pi}(\frac{\gamma}{2\omega_0})^3 + \cdots\;]\;.
\label{32}
\end{equation}
Notice that $\gamma/\omega_0 \approx 10^{-10}$ for an atomic oscillator.

Substituting the result (\ref{32}) in the expression (\ref{28}), we
get
\begin{equation}
\langle \overline{x^2} \rangle  =
\frac{\hbar}{m \omega_0}\; ,
\label{result}
\end{equation}
corresponding to a ground state energy that is {\it twice} the
correct value obtained by using the Heisenberg picture. As far we
know this discrepancy was first pointed out by A. J. Faria et al.
\cite{Faria}.

In the following section we shall show that neither the thermal
electromagnetic fields, nor the radiation reaction force, are
responsible for this discrepancy. It will be clear that the reason
for the discrepancy is that the zero-point fluctuations was
considered {\it twice} in the Schr\"{o}dinger equation. This was
suggested by Faria et al. in the section 5 of their paper
\cite{Faria}. We shall give the detailed proof that their suggestion
is correct.


\subsection*{2.2) The effects of the thermal electromagnetic fields and
  the radiation reaction in the Schr\"{o}dinger equation}

Our first observation refers to the momentum operator used in the
Schr\"{o}dinger equation (\ref{Sch}), namely, $p=-i\hbar
\frac{\partial }{\partial\ x}$. This operator already contains the
effects of the zero-point electromagnetic field. This can be easily
recognized from the works of P. W. Milonni \cite{Milonni,
Milonni81}. According to Milonni the commutator between the
operators $x(t)$ and $p(t)$ can be calculated within the Heisenberg
picture and the result is (see \cite{Milonni}, section 2.6)
\begin{equation}
[\; x(t),p(t)\;] = \frac{ie^2}{m} \frac{8 \pi}{3} \int_0^{\infty}
d\omega \frac{\omega \;\rho_0(\omega)}{(\omega^2 - \omega_0^2 )^2 +
\left(\frac{2e^2}{3mc^3}\omega^3 \right)^2} \;,
\label{comutador1}
\end{equation}
where $\rho_0(\omega)$ is the zero-point spectral density given
previously (see our equation (\ref{espectro})). Only $\rho_0(\omega)$
depends on $\hbar$.  The calculation of the integral (\ref{comutador1}) is similar to the calculation of
$\langle q_c^2(t)\rangle$ presented within the subsection 2.1. The
result is
\begin{equation}
[\; x(t),p(t)\;] = i\hbar\;.
\label{comutador}
\end{equation}
Noticed that the dipole approximation is used in order to obtain the
results (\ref{comutador1}) and (\ref{comutador}). The conclusion is
that, according to the Heisenberg picture, the constant $\hbar$
appearing in (\ref{comutador}) {\it has its origin in the zero-point
radiation with spectral distribution} $\rho_0(\omega) =
\hbar\omega^3/2\pi^2 c^3$. As we said above this was shown by P.W.
Milonni in the references \cite{Milonni, Milonni81}. In words of
Faria et.al. \cite{Faria}: ``the kinetic energy operator used in the
Schr\"{o}dinger picture, namely, $\frac{p^2}{2m} = -
\frac{\hbar^2}{2m} \frac{\partial^2 }{\partial\ x^2}$, is the
natural channel by means of which the zero-point electromagnetic
fluctuations are incorporated into the Schr\"{o}dinger equation
for a charged particle''.

Considering these observations, we conclude that {\it only} the
radiation reaction and the thermal electromagnetic fields of the
vacuum can be consistently introduced within the Schr\"{o}dinger
equation (\ref{Sch}). However, it is possible to replace the
zero-point field $E_0(t)$, used in the equation
(\ref{22}), by the thermal random field $E_T(t)$ given by
(\ref{ET}). We shall show that this replacement will lead to correct
results for several observable quantities. The discussion of the
effects of the thermal fields is an interesting and clarifying example.

Notice that the total electric field acting on the charged particle
will be $E_x(t) = E_T(t) + E_{RR}(t)$, where $E_{RR}(t)$ is the
radiation reaction field (see (\ref{24})). As before the vector
potential $A_x(t)$ is such that $E_x(t) = -
\frac{1}{c}\frac{\partial A_x}{\partial t}$.

Following the calculations explained within the subsection 2.1 (see
the equations (\ref{17}) to (\ref{25}), it is straightforward to
show that the function $q_c(t)$ will be given by the equation
(\ref{stat}), with the replacement of $\sqrt{\hbar\omega/2}$ by
$h(\omega,T)$ introduced in (\ref{hT}). Moreover, it is easy to show
that $\langle q_c^2(t) \rangle$ will be given by
\begin{equation}
\langle q_c^2(t) \rangle =
\frac{2 e^2}{3\pi m^2c^3} \int_0^{\infty}
d\omega \frac{\hbar\omega^3 \;[\;\coth(\frac{\hbar\omega}{2kT}) - 1\;]}{ ( \omega^2 - \omega_0^2 )^2 +
\left( \frac{2e^2}{3mc^3} \right)^2 \omega^6},
\label{34}
\end{equation}
instead of our previous equation (\ref{28b}). Notice that
$\coth(\frac{\hbar\omega}{2kT}) - 1 = (e^{\frac{\hbar\omega}{kT}} - 1)^{-1} $.

Introducing again the constant $\frac{\gamma}{\omega_0} =
\frac{2}{3}\frac{e^2}{\hbar c}\frac{\hbar \omega_0}{m c^2} \ll 1$,
the above integral can be calculated with the same approximations
used previously (subsection 2.1). With this procedure we get
\begin{eqnarray}
\langle q_c^2(t) \rangle &=& \frac{\hbar}{m\omega_0}\left(
\frac{1}{e^{\hbar\omega_0/kT}-1} \right)\left[1-\frac{1}{\pi}
\left(\frac{\gamma}{2\omega_0}\right)+\frac{1}{3\pi}\left(\frac{\gamma}
{2\omega_0}\right)^3+\cdots\right]\nonumber \\
& \simeq & \frac{\hbar}{m \omega_0}\left(\frac{1}{e^{\hbar\omega_0/kT}-1}\right)\;.
\label{35}
\end{eqnarray}
This new result combined with our previously expression (\ref{28})
gives
\begin{equation}
\langle \overline{x^2} \rangle = \frac{\hbar}{2m \omega_0}\left(1 +  \frac{2}{e^{\hbar\omega_0/kT}-1} \right)\;.
\label{36}
\end{equation}

This is the {\it correct} value of $\langle \overline{x^2} \rangle$ for an {\it arbitrary}
temperature $T$. Notice that we get $\langle \overline{x^2}
\rangle = \frac{\hbar}{2m \omega_0}$ \;when $T=0$. This is expected on
physical grounds, and is in agreement with calculation based on the
Heisenberg picture.

The conclusion is that the discrepancy between the Heisenberg and
the  Schr\"{o}dinger pictures, pointed out by A.J. Faria et al.
\cite{Faria}, is eliminated {\it only} when we {\it remove} the
zero-point field $E_0(t)$ from the Schr\"{o}dinger equation
(\ref{Sch}). {\it The effects of the zero-point field are already
included in the operator} $-i\hbar \partial/ \partial x$. The
inclusion of the fluctuating thermal electromagnetic fields into the
Schr\"{o}dinger equation is necessary and helps in the clarification
of this point.

We also would like to present another interesting effect of the
inclusion of the thermal electromagnetic fields (see (\ref{ET}) and
(\ref{hT})) into the Schr\"{o}dinger equation (\ref{Sch}). We recall
that (see (\ref{23}))
\begin{equation}
|\psi(x,t)|^2 =
\left( \frac{m\omega_0}{\pi\hbar} \right)^{\frac{1}{2}}
\exp{\left[ - \frac{m \omega_0}{\hbar}(x - q_c(t))^2 \right]}\;,
\end{equation}
where the fluctuating coordinate $q_c(t)$ depends on the temperature
$T$, and is such that
\begin{equation}
q_c(t) = \frac{e}{m} \sum_{\lambda=1}^{2} \int d^3k\;
\epsilon_x(\vec{k},\lambda) \frac{h(\omega,T)}{2\pi}
\left[\frac{e^{i\theta(\vec{k},\lambda)} e^{-i\omega t}
}{\omega_0^2-\omega^2-i\gamma\omega} + c.c. \right]\;. \label{qc}
\end{equation}

A quantity of interest is the temperature dependent probability density
\begin{equation}
P_T(x) \equiv \langle |\psi(x,t)|^2 \rangle\;,
\label{p0}
\end{equation}
where the symbol $\langle \; \rangle$ indicates average over the
random phases $\theta(\vec{k},\lambda)$.

In order to calculate (\ref{p0}) we shall introduce the Fourier transform
\begin{equation}
\int_{-\infty}^{+\infty}\frac{dk}{2\pi}e^{-\frac{\hbar
k^2}{4m\omega_0}}\;e^{-ik(x-q_c)} = \sqrt{\frac{m\omega_0}{\pi\hbar}}\;\exp[{-\frac{m\omega_0}{\hbar}(x-q_c(t))^2}].
\end{equation}
With this notation we obtain
\begin{equation}
P_T(x) = \int_{-\infty}^{+\infty}\frac{dk}{2\pi}e^{-\frac{\hbar
k^2}{4m\omega_0}}\;e^{-ikx} \langle e^{ikq_c(t)} \rangle.
\label{P}
\end{equation}

The calculation of the characteristic function $\langle
e^{ikq_c(t)} \rangle$ is standard. One can show that
\begin{equation}
\langle e^{ikq_c(t)} \rangle =
\sum_{n=0}^{\infty}\frac{(ikq_c(t))^n}{n!} = 1 - \frac{k^2\langle
q_c(t)^2\rangle}{2!} + \frac{k^4\langle q_c(t)^4\rangle}{4!} + \cdots\;,
\end{equation}
or
\begin{equation}
\langle e^{ikq_c(t)} \rangle = \sum_{n=0}^{\infty}(-1)^n\; \frac{k^{2n}\langle q_c(t)^{2n}\rangle}{(2n)!}\;.
\label{38}
\end{equation}
According to Boyer, (see eq. (68) of reference \cite{Boyer}), we have
\begin{equation}
\langle q_c(t)^{2n}\rangle =
\frac{(2n)!}{n!\;2^n}\left(\frac{\hbar[\coth
(\frac{\hbar\omega_0}{2kT}) - 1]}{2m\omega_0}\right)^n\;,
\label{39}
\end{equation}
where the factor $\left[ \coth(\frac{\hbar\omega_0}{2kT}) - 1 \right]$
has its origin in the expressions (\ref{qc}), for
$q_c(t)$, and (\ref{hT}) for $h(\omega,T)$.
Using (\ref{39}) and (\ref{38}) into (\ref{P}) we obtain
\begin{equation}
\langle|\psi(x,t)|^2\rangle =
\int_{-\infty}^{+\infty}\frac{dk}{2\pi}e^{-ikx}\exp\left[\;-\frac{\hbar k^2}{4m\omega_0}
\coth(\frac{\hbar\omega_0}{2kT})\;\right]\;.
\end{equation}
The integration is straightforward leading to the result
\begin{equation}
P_T(x) = \langle|\psi(x,t)|^2\rangle = \sqrt{\frac{m\omega_0}{\pi \hbar
\coth(\frac{\hbar\omega}{2kT})}} \exp\left[\;-\frac{m\omega_0 x^2}{\hbar \coth(\frac{\hbar\omega_0}{2kT})}\;\right]\;,
\label{probab}
\end{equation}
valid for an {\it arbitrary} temperature $T$. This result is observed experimentally.

Moreover, this expression for the quantum probability distribution
$P_T(x)$, associated with the probabilistic motion of a charged
harmonic oscillator, was derived here in a simple and direct manner.
It coincides with the result presented by R.W. Davies and K.T.R.
Davies \cite{Davies}. These authors have used the Wigner phase space
distribution functions associated with the discrete excited states
of the harmonic oscillator, obtained according to the
Schr\"{o}dinger picture. In their work, the temperature effects are
introduced by the use of the Boltzmann factors associated with each
excited state. Notice that Davies and Davies \cite{Davies} do not
mention, neither the effects of the thermal electromagnetic fields
(see (\ref{ET})), nor the dynamical role of the radiation reaction
force. In this respect our calculation elucidates in details, and
more clearly, the influence of the radiation reaction and the
radiation bath on the oscillator.


\section*{3. Conclusions}

The SED approach is mainly used in the study of linear
systems or other systems that can be treated linearly in a good
approximation. Interesting examples are the interaction of electric
and magnetic dipoles with simple circuits with thermal and
zero-point voltage fluctuations, as the RLC circuit. New findings
were obtained in this way, and were published recently \cite{Blanco,
Santos1, Santos2, Noise, Noise2}. We call the reader attention to
the work of Blanco et.al. \cite{Noise} concerning the enhancement of
the voltage fluctuations by the action of the classical zero-point
magnetic field in the coils of an appropriated constructed solenoid.
This new prediction of SED is currently under experimental
investigation by L.J. Nickisch \cite{LJ}. A nonlinear phenomenon,
namely, the "tunneling" from a potential well with a barrier, was 
successfully explained as an effect of the zero-point radiation \cite{Tunneling}.

In our paper we have studied a fundamental problem which is
possible to be treated within the realm of SED. We have extended the
analysis of A. J. Faria et al. \cite{Faria}, by considering the
effects of the {\it zero-point} and {\it thermal} electromagnetic
fields in a harmonic oscillator using the {\it Schr\"{o}dinger
picture}. The effects of the radiation reaction were also correctly
taken into account. We concluded that {\it the effects of the
zero-point radiation are already included into the Schr\"{o}dinger equation
(\ref{Sch}) by means of the momentum operator}\;
$p=-i\hbar\frac{\partial}{\partial x}$ (see also the references
\cite{Faria, Dalibard,Milonni81} which complement this statement).
In our opinion, this is the most important finding of our work.
Such a conclusion stresses the problem of the {\it equivalence} between
the Schr\"{o}dinger and the Heisenberg pictures of Quantum Mechanics
(see the references \cite{Dirac, Faria}) as far the harmonic
oscillator is concerned. This is easy to understand because the
effects of the zero-point field (\ref{E0}) has to be {\it
subtracted}, in the Schr\"{o}dinger treatment of the oscillator, in
order to obtain the correct result (see the eq. (\ref{result}) in the
subsection 2.1). This {\it subtraction is not
necessary in the case of the Heisenberg picture} \cite{Faria,
Milonni}.

\section*{Acknowledgements}
One of us (H.M.F.) wants to thank Professor Jean-Pierre Vigier for
interesting comments concerning the subject of this paper, and to
Prof. Coraci P. Malta for a critical reading of the manuscript.
We acknowledge the financial support from Funda\c c\~ao de Amparo \`a
Pesquisa do Estado de S\~ao Paulo (FAPESP) and Conselho Nacional de
Desenvolvimento Cient\'{\i}fico e Tecnol\'ogico (CNPq - Brazil).

\end{document}